\newcommand{\lya}{Ly$\alpha$}
\newcommand{\halpha}{H$\alpha$}
\newcommand{\hbeta}{H$\beta$}
\newcommand{\hi}{H{\sc i}}
\newcommand{\heii}{He{\sc ii}}
\newcommand{\oii}{[O{\sc ii}]}
\newcommand{\oiii}{[O{\sc iii}]}
\newcommand{\nii}{[N{\sc ii}]}
\newcommand{\fnu}{$f_\nu$}
\newcommand{\fluxcgs}{erg~s$^{-1}$~cm$^{-2}$}

\newcommand{\Msunyr}{$M_\odot$~yr$^{-1}$}

\newcommand{\ebv}{$E_{B-V}$}
\def\aap{A\&A}
\def\aj{AJ}
\def\apj{ApJ}
\def\apjl{ApJL}
\def\apjs{ApJS}
\def\nat{Nature}
\def\araa{ARA\&A}
\def\mnras{MNRAS}

\documentclass[useAMS,usenatbib]{mn2e}
\usepackage{amssymb,amsmath}
\usepackage{natbib,graphicx}
\usepackage{color}

\title[A peculiar redshift 11 interloper]
{A peculiar galaxy appears at redshift 11: properties of a moderate redshift interloper\thanks{Based on observations made with ESO Telescopes at the Paranal Observatory under programme ID 087.A-0118}}
\author[Hayes et al.]{Matthew Hayes$^{1,2}$\thanks{E-mail: matthew.hayes@irap.omp.eu (MJH)}, 
Nicolas Laporte$^{1,2}$, 
Roser Pell\'o$^{1,2}$, 
Daniel Schaerer$^{3,2}$, \and 
Jean-Fran\c{c}ois Le Borgne$^{1,2}$\\
$^{1}$Universit\'e de Toulouse; UPS-OMP; IRAP; Toulouse, France\\
$^{2}$CNRS; IRAP; 14, avenue Edouard Belin, F-31400 Toulouse, France\\
$^{3}$Observatoire de Gen\`eve, Universit\'e de Gen\`eve, 51 Ch. des Maillettes, 1290 Versoix, Switzerland
}
\begin{document}

\date{Accepted 2012 May 29. Received 2012 May 17; in original form 2012 April 13}

\pagerange{\pageref{firstpage}--\pageref{lastpage}} \pubyear{2002}

\maketitle

\label{firstpage}

\begin{abstract}
\citet{Laporte2011} reported a very high redshift galaxy candidate: a lensed 
$J-$band dropout (A2667-J1). J1 has a photometric 
redshift of $z=9.6-12$, the probability density function for which permits no 
low or intermediate $z$ solution.  We here report new spectroscopic 
observations of this galaxy with VLT/XShooter, which show clear 
\oiii$\lambda 5007$\AA, \lya, \halpha, and \hbeta\ emission and 
place the galaxy firmly at $z=2.082$. 
The oxygen lines contribute only $\sim 25$\% to the $H-$ band flux, 
and do not significantly affect the dropout selection of J1. After 
correcting the broadband fluxes for line emission, we identify two roughly
equally plausible natures for A2667-J1: either it is
young heavily reddened starburst, or a maximally old system with a
very pronounced 4000\AA\ break, upon which a minor secondary burst of star 
formation is superimposed. Fits show that to make a 
$3\sigma$ detection of this object in the $B-$band ($V-$band), imaging of 
depth AB=30.2 (29.5) would be required -- despite the relatively 
bright NIR magnitude, we would need optical data of equivalent 
depth to the Hubble Ultra Deep Field to rule out the mid-$z$ solution
on purely photometric grounds. Assuming that this stellar population can be 
scaled to the NIR magnitudes of recent HST/WFC3 IR-selected galaxies, we 
conclude that infeasibly deep optical data AB$\sim 32$ would be required for 
the same level of security. There is a population of galaxies at $z\approx 2$
with continuum colours alone that mimic those of our  $z = 7-12$ 
candidates.

\end{abstract}


\section{Introduction}
Broadband imaging in the near infrared (NIR) has enabled the detection
of candidate galaxies at redshifts between $z=7$ and 12, using both recent
survey instruments on the Hubble Space Telescope 
\citep{McLure2010,Finkelstein2010,Bouwens2011a,Trenti2011,Yan2011} 
and wider--but--shallower observations 
from the ground (E.g. \citealt{Ouchi2009,Capak2011}, Pell\'o et al in prep). With 
specific astrophysics in mind, these catalogues may be used to infer the cosmic history 
of star formation, the photon budget available for reionization, 
and also as an independent test of the re-ionization edge through the prevalence
of the H{\sc i} \lya\ emission line 
\citep{Pentericci2011,Schenker2012,Curtis-Lake2012}. 

It is vital, therefore, to understand the lower-$z$ objects that could contaminate 
such samples,
and the frequency with which they do. At the highest $z$, contamination
rates have been estimated using population synthesis methods
\citep{Trenti2011,Bouwens2011a}, and shown to be non-negligible. Empirical studies
at $z\sim 2$ have also found a population of galaxies with emission lines 
strong enough to mimic a strong spectral break,
suggesting roughly similar contamination levels \citep{Atek2011}.
Yet the empirical interloper rate, measured directly by deep spectroscopy in the 
$z>7$ samples, remains broadly unknown.

To these ends our group has been carrying out its own multi-wavelength studies, 
taking advantage of gravitational lensing to search for intrinsically fainter 
background objects \citep[][ L11]{Laporte2011}
Originally, L11 selected ten 
lensed galaxies that satisfy criteria for $z-$, $Y-$, and $J-$dropout galaxies. 
With deep optical, near-infrared and Spitzer/IRAC photometry, traditional 
photometric redshift (phot-$z$) methods placed all galaxies in the
range $z=7-12$. 
However two of the L11 sources were found to lie at $z\sim 2$ 
\citep{Boone2011},
hinting that strong optical lines may have mimicked a Lyman break. 
In cases where very deep optical imaging is available, this source of contamination 
is thought to be small, as the line strengths imply lowish masses/metallicities
that in turn should exhibit blue continuum colours and remain detectable, 
particularly when stacked.  However other sources of contamination 
exist, specifically relating to dust attenuation and strong 4000\AA\ breaks: in 
both of these cases an intrinsically red spectrum in the restframe UV and 
optical may cause a galaxy to go undetected in all bandpasses bluer than the 
first dropout band. 

In this Letter we present the optical and NIR spectrum of a bright 
$J-$band dropout galaxy (A2667-J1)  -- phot-$z=11.9^{+>0.1}_{-2.3}$, no 
low-$z$ solution and no Herschel detection -- obtained with VLT/XShooter. 
In L11 we were skeptical about the very high-$z$ interpretation because 
of the large implied luminosity, and the detection of a
faint source in HST/$z_{850}$ imaging (although the $z-H$ colour still exceeds $2$ magnitudes).
However the spectrum shows several 
strong emission lines that are consistent only with $z=2.082$, implying
it is an interloper. 
In Sect.~\ref{sect:data} we present the new observations and measurements of emission lines.
In Sect.~\ref{sect:whatis} we discuss inferences we can draw from the emission lines, 
and explanations for the strong spectral break and selection function, and re-model the
stellar continuum. In Sect.~\ref{sect:implic}
we discuss the implications that this galaxy may have upon other very high-$z$ surveys. 
All magnitudes are AB. 

\section[]{Observations and Measurements}\label{sect:data}

\subsection[]{X-Shooter observations }

A2667-J1 ($\alpha$=23:51:34.9, $\delta$=--26:3:32.7, J2000) was observed 
with X-Shooter \citep{DOdorico2006,Vernet2011},
mounted at UT2 of ESO/VLT, on the night of 24 August 2011. The total 
integration time was short, but at 
$4\times 900$~s was sufficient to find several strong emission lines. 
We used the 1\arcsec\  (0.9\arcsec) slit for the UVB channel (VIS and NIR),
providing an almost constant aperture between the $U-$ and $K-$ bands. More 
importantly the NIR spectral resolution of $R>5000$ 
is sufficient to resolve the atmospheric OH airglow lines.
Between each observation, the object was dithered along the slit 
by 2.5 \arcsec. The average atmospheric seeing was 0.76 \arcsec. 
Data were reduced and combined using the {\sc EsoRex} X-Shooter pipeline version 1.3.7, and 
one dimensional spectra were extracted from the frames using the {\sc apall} task in 
{\sc noao/iraf} with an aperture width of 1.5 \arcsec.

\subsection[]{Emission lines in the spectrum}

\begin{figure*}
 \includegraphics[width=170mm]{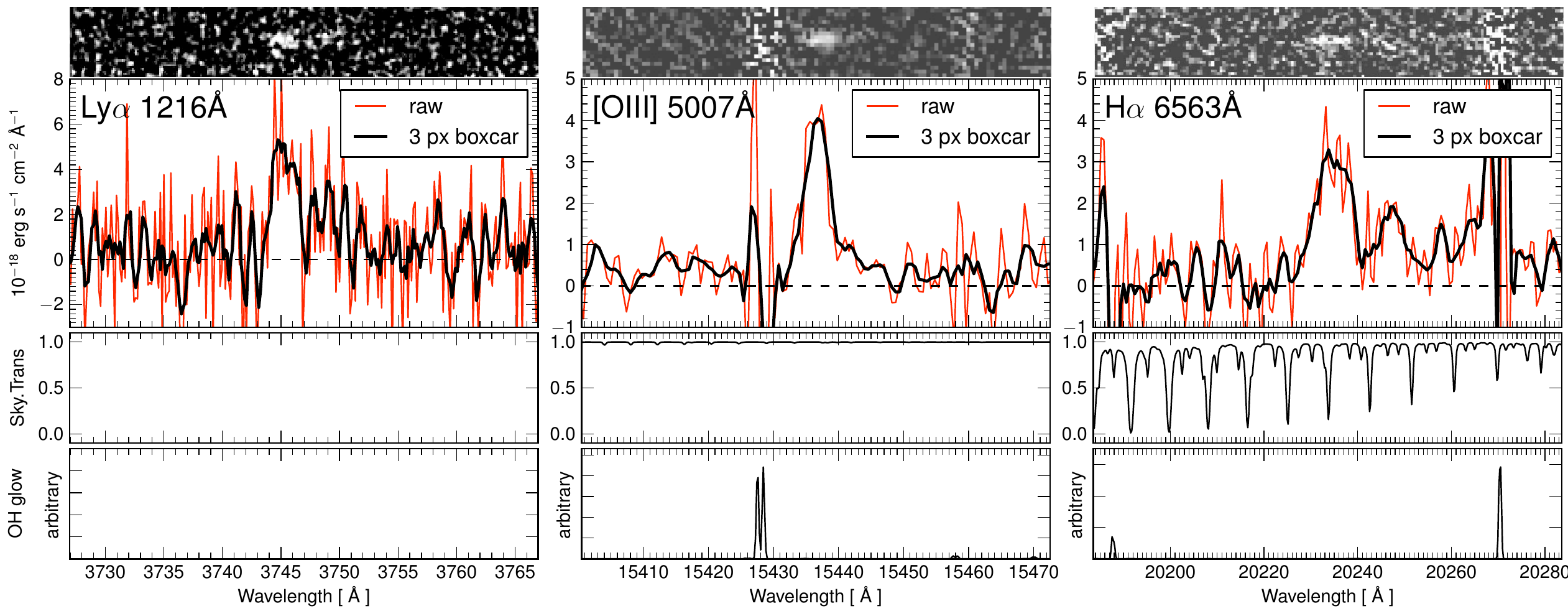}
 \caption{Examples of the extracted 1D spectra around strong detections. \emph{Left} shows
	\hi\ \lya, \emph{centre} shows \oiii$\lambda 5007$\AA, and \emph{right} shows \hi\ \halpha.
	From \emph{top}	to \emph{bottom} each panel shows the 2D spectral images, the 1D spectral
	extractions, the sky transmissivity/telluric absorption, and the OH airglow 
	spectrum. \lya\ presents with its characteristic asymmetric shape. \oiii\ is the strongest
	line, and detected with $S/N=20$, while \halpha\ appears somewhat contaminated by both
	telluric bands and skylines.  See text for details.  }
 \label{fig:spectra}
\end{figure*}

We clearly identified a strong emission line at wavelength
$\lambda=15437$\AA\ (Fig.~\ref{fig:spectra}) -- 
after masking regions contaminated by a nearby airglow line 
and fitting a Gaussian profile, we measure a flux of 
$(18.8\pm0.96) \times 10^{-18}$~\fluxcgs. 
The $H-$band magnitude is $25.21 \pm 0.08$ (L11), and assuming the quoted
bandpass parameters ($\lambda_\mathrm{c}=1.620$\micron, FWHM=2890\AA), we 
compute an observer frame equivalent width, $EW$, of $(679\pm 71)$\AA. The 
wavelength of the  observed emission line falls close to the peak transmission 
so no filter wing corrections are necessary. 

This high $EW$  restricts the possible line identifications - after testing various possibilities, 
we determine it to be \oiii$\lambda 5007$\AA\ 
and also identify \lya, \hbeta, and \halpha. This places
the redshift of J1 firmly at $z=2.082$. After the removal of a nearby OH line
\oiii$\lambda 4959$\AA\ is also visible but no \heii$\lambda 1640$\AA, 
\nii$\lambda\lambda$6548,6584, or 
\oii$\lambda\lambda$3726,3729\AA\ are visible in either the 1D or 2D spectra. 
Example spectra are shown in Fig.~\ref{fig:spectra}, and the measured line fluxes (or limits) 
are listed in Table~\ref{tab:fluxes}. 

\begin{table}
 \begin{minipage}{80mm}
  \caption{A2667-J1 fluxes in emission lines. \label{tab:fluxes}}
  \begin{tabular}{@{}lllll@{}}
  \hline
   Species & $\lambda_\mathrm{rest}$ & $\lambda_\mathrm{meas}$ & $z$ & Flux \\
           &  \AA\        &  \AA\            &          & $10^{-18}$ \fluxcgs\ \\
 \hline
 \hi\ \lya    & 1215.67 & 3743.0  & 2.079 & $ 9.41 \pm  2.25 $ \\
 \heii\       & 1640.42 & ...     & ...   & $ ~~~~~ <4.87 $ \\
 \oii\        & 3727.09 & ...     & ...   & $ ~~~~~ <2.72 $ \\
 \oii\        & 3728.79 & ...     & ...   & $ ~~~~~ <2.72 $ \\
 \hi\ \hbeta  & 4862.72 & 14998.2 & 2.084 & $ 7.77 \pm  4.22 $ \\
 \oiii\       & 4960.30 & 15288.2 & 2.082 & $ 7.06 \pm  2.72 $ \\
 \oiii\       & 5008.24 & 15436.6 & 2.082 & $ 18.8 \pm  0.96 $ \\
 \hi\ \halpha & 6564.61 & 20234.6 & 2.082 & $ 22.7 \pm  4.08 $ \\
 \nii\        & 6583.46 & 20290.2 & ...   & $ ~~~~~ <3.62 $ \\
\hline
\end{tabular}
\end{minipage}
\end{table}

\section[]{What is A2667-J1?}\label{sect:whatis}

\subsection{Basic inferences from the nebular lines }\label{sect:whatislines}

We first discuss a few line ratios. 
Both \halpha\ and \hbeta\ can be measured, although the 
significance of \hbeta\ is low $(2\sigma)$. The \halpha/\hbeta\ ratio is $2.92 \pm 1.67$, 
consistent with the unobscured value of 2.86 ($T=10^4$~K), which corresponds to 
\ebv$=0\pm0.5$ ($1\sigma$,  \citealt{Calzetti2000}). 
Assuming no obscuration, the star formation rate (SFR) inferred from \halpha\ is $4.1\pm 0.7$\Msunyr\ 
(\citealt{Kennicutt1998}, corrected for a magnification), which 
is towards the lower end of the envelope of \citet{Hayes2010a} for $z\sim 2$ \halpha-selected galaxies.
At the $1\sigma$ upper limit of dust 
attenuation the intrinsic SFR will be a factor of 5 higher. 
Regarding \lya, assuming pure recombination under case B  and 
the permitted range of \ebv, \lya\ escape fractions falls in the range 1--5\%, again close to the 
$z\approx 2$ average \citep{Hayes2010b}. 

Neither side of the \oii\ doublet is detected ($<1.92$~\fluxcgs), despite the region being 
free of both telluric and airglow features.
This allows us to put a reasonable limit on the excitation 
parameter, $P$ [=\oiii/(\oiii+\oii)] of $1.0^{+0}_{-0.1}$, which is not greatly affected by the range
of permitted dust contents. $P$ is known to correlate strongly with the ionization parameter, 
and the hardness of the ionizing spectrum (therefore the temperature of the stellar 
population, E.g. \citealt{Pilyugin2001}), and the spectrum suggests that the stellar population of 
A2667-J1 is still hot enough to keep all the oxygen doubly ionized. We may also calculate the 
$R_{23}$ index [(\oii+\oiii)/\hbeta] which is not particularly high at $3.2\pm 1.6$; the high 
value  of $P$, the low $R_{23}$, and the absence of a clear \nii\ lines surrounding \halpha\ suggest 
that we are looking at a relatively  low metallicity burst of star formation. 
From the N2 index, metallicity $12+\log(\mathrm{O/H}) < 8.39$ 
($1\sigma$, \citealt{PettiniPagel2004}), which would place J1 roughly midway 
through the distribution of \citet{Erb2006} at $z\sim 2$. While J1 is very compact, 
the \oiii/\hbeta\ ratio and \nii/\halpha\ limit are fully consistent with a nebula produced
by 	star-formation, not an active nucleus \citep{Kauffmann2003}.

\subsection[]{Line contribution and colour selection }

A2667-J1 met the selection criteria for a $z\sim 11$ galaxy in [$(J-H)$ vs.
$(H-K_\mathrm{s})$] colour space defined in L11 and optimized
for the removal of stars and lower $z$ galaxies. It is 
important to note that every survey has its own combination of wavelength 
coverage and depth, and therefore the selection criteria will naturally differ.
Indeed most comparable studies are based upon HST data, and consequently include
no $K-$band criteria for their $J-$band dropouts. Details aside, it is well known 
that \oiii\ and \halpha\ lines may have $EW$ high enough to dominate broadband fluxes 
\citep[E.g.][]{SchaererDeBarros2009,Trump2011} causing mid-$z$ galaxies to 
interlope into very high-$z$ samples  \citep{Atek2011}. A vital question,
therefore, is whether J1 was selected because of this. 
Combined \oiii+\hbeta\ provide $EW \approx 900$~\AA, and these lines 
contribute just 25\% to the $H-$band flux ($FWHM=2890$~\AA). If we subtract 
this nebular contribution, the colour of the $(J-H)$ spectral break changes 
from the measured value of $>2.3$ magnitudes to $>2.0$, and still easily clears
the dropout criterion of $(J-H)>0.76$. After subtracting 
\halpha\ from the $K_\mathrm{s}$ band we recompute the 
$(H-K_\mathrm{s})$ colour as 0.50, only slightly redder than the 0.44 
measured in L11 and still on the border of the selection box. 
Therefore even without the strong nebular lines we would 
still have selected J1 as a very high-$z$ candidate. 

There is the need to explain a spectral discontinuity with amplitude $>6$ in 
\fnu\ between the $J-$ and $H-$bands (restframe 5250 and 4080~\AA), and 
amplitude $>10$ between $Y$ (restframe 3310~\AA) and $H$. 
This could result from either heavy dust obscuration or a 
strong 4000\AA\ break -- the former seems inconsistent with \halpha/\hbeta, while 
the latter  seems inconsistent with even the presence of strong emission lines.

\subsection[]{(Re-) Modeling the stellar population}\label{sect:whatismodel}

\begin{figure*}
 \includegraphics[width=88mm]{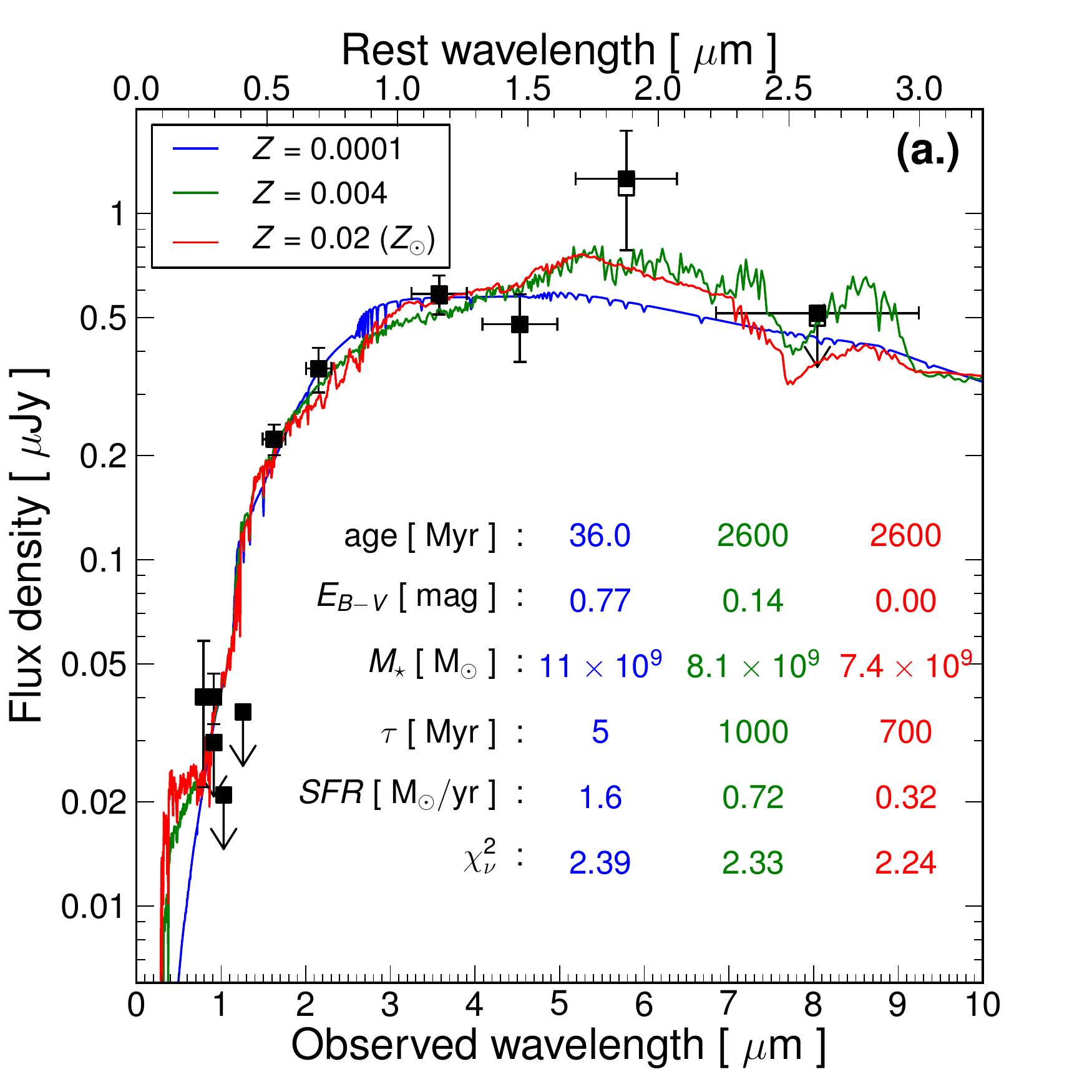}
 \includegraphics[width=88mm]{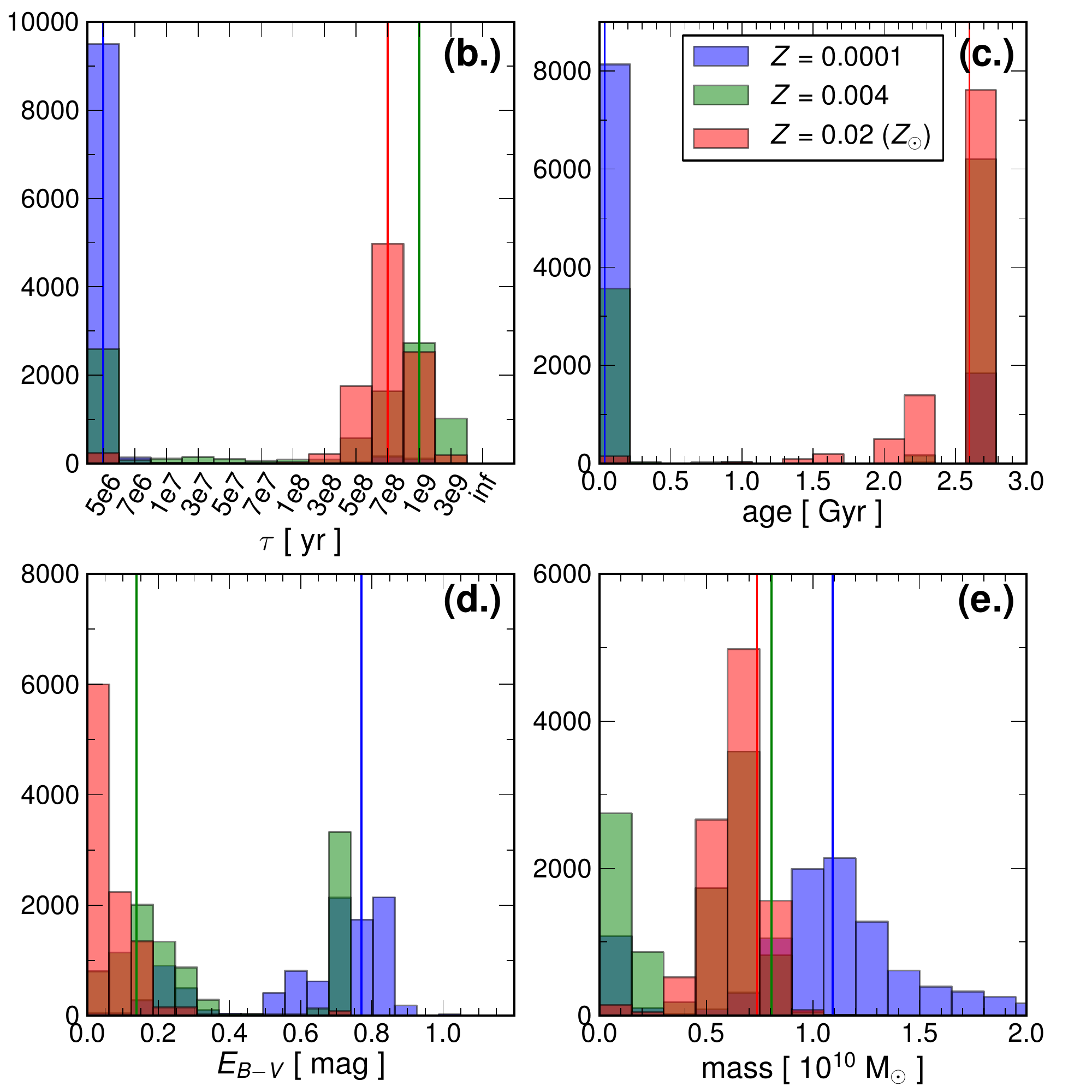}
 \caption{SED fitting results using \emph{Hyper-z} and the CB2007 synthesis models.
Panel a. (\emph{left}) shows the fits, with data points and limits as solid black boxes; error bars are 
$1\sigma$ and horizontal bars represent the FWHM of the filters. Open boxes show the IRAC points 
when corrected for Paschen and Brackett line emission. The solid coloured lines show different 
stellar metallicities; inferred quantities are displayed in the corresponding colour. 
The \emph{right} panel shows the distribution of the quantities recovered from fits to $10^4$ 
Monte Carlo realization of the data. The same metallicity colour scheme applies. 
We show SFH $\tau$ (b.), age (c.), \ebv\ (d.), 
	 and stellar mass (e.).}
 \label{fig:fits}
\end{figure*}

Before modeling, we take advantage of new Spitzer data reduction 
methods and re-perform the IRAC photometry using the latest pipeline processed
data. The high-level science data from the Lensing Cluster GTO Survey 
(Program 83, PI. Rieke) were re-obtained, and photometry 
performed in 1~arcsec circular apertures paying special attention to the 
sky properties (level, noise, skewness) in nearby empty 
apertures. In Table~\ref{tab:newphot} we present updated stellar (i.e. 
line-subtracted) VLT photometry, updated IRAC photometry, and the original 
HST photometry from L11. We note that in the optical and NIR colours, A2667-J1 
appears rather similar to the unusual infrared object HDFN-JD1 
\citep{Dickinson2000}, although it is 0.7 magnitudes bluer in $H-K_\mathrm{s}$ --
the nature of that source also remains unclear. 

\begin{table}
 \begin{minipage}{80mm}
  \caption{Updated stellar fluxes, corrected for emission lines\label{tab:newphot}}
  \begin{tabular}{@{}lllll@{}}
  \hline
   Telescope   & obs $\lambda_\mathrm{c}$ & rest $\lambda_\mathrm{c}$ & magnitude  \\
   \& bandpass & [ \micron\ ]      &  [ \micron\ ]              & [ AB ] \\
   \hline
   \emph{VLT/FORS2/I}              & 0.791 & 0.257 &  $~~~27.39 \pm 0.49$\footnote{as reported in L11} \\
   \emph{HST/ACS/F850LP}           & 0.911 & 0.296 &  $~~~27.39 \pm 0.18$ \\
   \emph{VLT/FORS2/z}              & 0.914 & 0.297 &  $>27.72^a$ \\
   \emph{VLT/HAWKI/Y}              & 1.03  & 0.334 &  $>28.1^a$          \\
   \emph{VLT/HAWKI/J}              & 1.26  & 0.409 &  $>27.5^a$ \\
   \emph{VLT/HAWKI/H}              & 1.62  & 0.526 &  $~~~25.53 \pm 0.11$\footnote{emission line subtracted} \\
   \emph{VLT/HAWKI/K$_\mathrm{s}$} & 2.15  & 0.698 &  $~~~25.02 \pm 0.16^b$ \\
   \emph{Spitzer/IRAC/Ch1}         & 3.58  & 1.16  &  $~~~24.48 \pm 0.14$\footnote{photometry modified from L11} \\
   \emph{Spitzer/IRAC/Ch2}         & 4.53  & 1.47  &  $~~~24.70 \pm 0.24^c$ \\
   \emph{Spitzer/IRAC/Ch3}         & 5.78  & 1.88  &  $~~~23.65 \pm 0.41^c$ \\
   \emph{Spitzer/IRAC/Ch4}         & 8.05  & 2.61  &  $>24.62^c         $ \\
\hline
\end{tabular}
\end{minipage}
\end{table}

With new stellar (i.e. line-corrected) photometry, we 
model the stellar population with \emph{Hyper-z} \citep{Bolzonella2000},
using the CB2007 modification of the \citet{BruzualCharlot2003} population 
synthesis models to generate evolutionary templates. We adopt a Salpeter
IMF, lock the redshift $z=2.082$, and fit dust attenuation, star formation
history (SFH), and time since the onset of star formation. The range of included
SFHs include exponentially decaying SFRs with $e$-folding timescales as 
listed in the abscissa of Fig~\ref{fig:fits}, panel (b). For comparative purposes
we perform fits at three different metallicities: 
$Z=0.0001$, 0.004, and 0.02. We show the results in the left panel of Fig.~\ref{fig:fits}, with 
histograms of $10^4$ Monte Carlo realizations in the right panel. 

Firstly we note that for the three different metallicities, the values of 
$\chi^2$ that describe the best fit are effectively consistent, and we do not have the power
to distinguish between them. Metallicity does however modulate the frequency
with which other results are obtained, and there are two 
different and entirely degenerate ways of reproducing the break: either with a
large dust content or an old population with a strong 4000\AA\ break. The latter
scenario would imply a dual stellar population: one evolved enough 
for a strong break, the other young enough to produce the emission lines. 
As usual 
a strong 4000\AA\ break is more frequently seen at higher metallicities, where
stellar evolution runs faster. Indeed models at extremely low $Z$ prefer the
young dusty solution in $>80$\% of cases, models at $Z=0.004$ seem to be rather 
evenly split between the young/dusty vs. old solutions, and the $Z=0.02$
models almost exclusively prefer ages above 2~Gyr and little dust.
The $Z=0.0001$ models are rejected by three arguments: (\emph{i.})
the necessary attenuation is higher than permitted from \halpha/\hbeta\ 
(Sect~\ref{sect:whatislines}), while  usually nebular \ebv\ exceeds that 
of the stars \citep{Calzetti2000};
(\emph{ii.}) such metallicities should produce higher \oiii/\hbeta;
and (\emph{iii.}) a dusty metal-poor galaxy seems an unlikely combination. However
the $Z=0.004$ models all seem permissible, and we do not have the information
to formally distinguish between the two scenarios.

We note that in the case of the composite population, the recent burst must 
also have a stellar population that has not been subtracted from the photometry
prior to fitting. Indeed its contribution 
to the $J-$band flux is not permitted to be large, otherwise the 
$J-H$ break would be diminished. To investigate this we take 
the restframe \halpha\ EW (280\AA) and 
compute the range of ages at which this EW is exceeded:
the star-forming population must have an age below $\approx 5$Myr 
\citep[][\emph{Starburst99}]{Leitherer1999}. Over this age range we then 
compute the range of permitted $J$ magnitudes and without reddening the
population will remain completely
undetected in $J$ for ages below 2~Myr. However the \halpha\
EW is independent
of reddening while the $J-K_\mathrm{s}$ colour is not; the $1\sigma$ limit of 
\ebv\ allows up to a factor of 3 greater attenuation at $J$ than at $K$ and ages 
up to 4~Myr still predict a $J-$band non-detection. 
On the other hand, a rather blue continuum may explain the HST $z_{850}$
detection.  For 
this range in parameter space, the burst contributes between 0.4 and 4~\% of the 
total stellar mass. 

Although the IRAC channels show the largest errors, the 
5.8\micron\ channel carries a residual of $\approx 1.2\sigma$. We note 
that the higher metallicity stars show the strongest 1.6\micron\ stellar 
feature, which is probably the cause of the marginal preference for these models.
However the IRAC points also contain the Paschen and Brackett lines, for which
we also make a first order attempt to correct (assuming $10^4$K gas). Since the 
nebular extinction is not well constrained, we make two corrections, one derived 
for \ebv=0 and the other for \ebv=0.5. These modified points are shown in 
Fig~\ref{fig:fits} and the largest correction (where Pa$\alpha$ falls in 5.8\micron)
amounts to just 0.06 magnitudes - consequently the corrections in channels 1 and 2 
are not visible. We do not re-perform the SED fit, but note that this correction 
does just bring the channel 3 data point to about $1\sigma$ of the fit.

Finally we examine the phot-$z$ solutions 
that are allowed by the new line-free photometry, and we repeat 
the fits with the redshift free. The peak at $z=10-12$ is still recovered but 
now represents the tertiary solution. The primary solution lies at $z=2.6-3.3$ 
which is actually in good agreement with the secondary peak from 
L11 ($z=2.8$), and the secondary grouping prefers $z<1.5$.
However, more interestingly, only 0.2\% of the realizations have phot-$z$ within
$1\sigma$ of the spectroscopically measured value, and it is not easy to 
obtain a good SED fit at the measured redshift when 
the templates comprise only single stellar populations. This lends support to the 
scenario of multiple populations, and the hypothesis that the $z_{850}$ and NIR
detections do not see the same stars. 

\section{Implications for high-z surveys}\label{sect:implic}
A2667-J1 can be described in two ways, but
regardless of the true population, one salient fact remains: 
was photometrically selected as a $z\sim 11$ candidate, but 
actually lies at $z\approx 2$. 

Because of our homogeneous coverage across the NIR, J1 was selected 
in two colour space: ($J-H$) and ($H-K_\mathrm{s}$). The selection, therefore,
is more similar to lower $z$ methods (e.g. $z-$ and $Y-$dropouts) and somewhat
different from other $J-$drop searches \citep{Henry2007,Bouwens2011b}, which lack 
comparably deep $K-$band data and therefore employ a one-colour 
selection. The requirement of $(H-K_\mathrm{s})<0.5$ (a blue UV slope) 
is specifically intended to fortify the selection against the inclusion of 
mid-$z$ galaxies with strong optical breaks, but a $z\approx 2$ galaxy is 
precisely what we have found. On the other hand, a non-detection in bandpasses 
bluewards of the break is also required by typical surveys to reject lower-$z$ 
galaxies and, while J1 is undetected in ground-based $z$, $Y$, and $J$ imaging, it
is detected in HST $z_{850}$ (albeit with a colour similar to the dropout criterion).
For J1, imaging bluer than the $z$ band would be effectively useless
in evaluating its status as an interloper: from the best-fitting SEDs and Monte Carlo
simulations described in Sect.~\ref{sect:whatismodel}, we calculate that in the 
HST F435W and F606W ($B$ and $V$) we would need to reach depths of AB=30.2 and 29.5 
in order to detect the continuum. In other words, this relatively bright $H-$detected
$J-$drop would need imaging with  the depth of the Hubble Ultra Deep 
Field \citep{Beckwith2006} to be rejected as very high-$z$.
A magnitude $H=28$ galaxy of these colours in the HUDF or CANDELS surveys would still remain 
undetected  by in the available HST $z-$band imaging, and require imaging reaching $V=32$ 
in order to detect the bluer continuum. It currently remains unclear 
how abundant galaxies of this type are, and thus systematic campaigns of spectroscopic 
followup will remain a vital pursuit, even at the faintest magnitudes. 

\section*{Acknowledgments}

This work received support from Agence Nationale de la recherche bearing the 
reference ANR-09-BLAN-0234-01. D.S. acknowledges support from the Swiss \
National Science Foundation.

\bibliographystyle{mn2e}

\bsp

\label{lastpage}

\end{document}